\documentclass[epj]{svjour}
\usepackage{graphics}
\begin{document}
\title{Bounds on the presence of quantum chaos in nuclear masses}
\author{Jorge G. Hirsch\inst{1}, 
   Alejandro Frank\inst{1},
   Jos\'e Barea\inst{1} \and
   Piet Van Isacker\inst{2} \and
 V\'\i ctor Vel\'azquez\inst{3}
\thanks{This work was supported in part by the Conacyt, Mexico, and DGAPA-UNAM.}%
}                     
%
%
\institute{Instituto de Ciencias Nucleares,
Universidad Nacional Aut\'o\-noma de M\'exico,
AP 70-543, 04510 M\'exico DF, Mexico 
\and 
GANIL,
BP 55027, F-14076 Caen Cedex 5, France
\and
Departamento de F\'\i sica, Facultad de Ciencias,
Universidad Nacional Aut\'o\-noma de M\'exico,
AP 70-348, 04511 M\'exico DF, Mexico}
%
%
\abstract{
Differences between measured nuclear masses and those calculated using the Finite Range Droplet Model are analyzed. It is shown that they have a well defined, clearly correlated
oscillatory component as a function of the proton and neutron numbers. At the same time,
they exhibit in their power spectrum the presence of chaos. Comparison with other mass calculations strongly suggest that this chaotic component arises from many body effects not included in the mass formula, and that they do not impose limits in the precision of mass calculations.
\PACS{21.10.Dr, 05.40.-a, 24.60.Lz, 05.45.Tp } 
} 
\maketitle
\section{Introduction}
\label{intro}

It has been recently proposed that there might be an inherent limit to the accuracy with which nuclear masses can be calculated \cite{Abe02}, due to the presence of chaotic motion inside the atomic nucleus \cite{Boh02}. This suggestion could have important consequences in the fields of nuclear physics and astrophysics, because
the knowledge of nuclear masses is of fundamental importance for
a complete understanding of the nuclear processes that power the
Sun and for the synthesis and relative abundances of the elements
\cite{Rol88}. 

Though great progress has been made in the challenging task of measuring the mass of exotic nuclei, theoretical models are necessary to {\em predict} their mass
in regions far from stability \cite{Lunn03}.
The simplest one is that of the liquid drop model (LDM). It incorporates the essential macroscopic terms, which means that the nucleus is pictured as a very dense, charged liquid drop.
The finite range droplet model (FRDM) \cite{Moll95}, which combines the macroscopic effects
with microscopic shell and pairing corrections, has become the {\em de facto} standard for mass formulas.
A microscopically inspired model has been introduced by Duflo and Zuker (DZ) \cite{Duf94}
with good results.
Finally, among the mean-field methods it is also worth mentioning
the Skyrme-Hartree-Fock approach \cite{Gor01}.

Besides the ``global" formulas of which the FDRM method has become the standard, there are a number of ``local" mass formulas. These local methods are usually effective when we require the calculation of the mass of a nucleus, or a set of nuclei, which are fairly close to a number of other nuclei of known mass, exploiting the relative smoothness of the masses M(Z,N) as a function of proton (Z) and neutron (N) numbers to deduce systematic trends.  Among these methods there are a set of algebraic relations for neighboring nuclei, known as the Garvey-Kelson (GK) relations \cite{Gar66}.

These relations do not have any free parameters and can be  derived from an independent particle picture.  They are based on a clever idea. The combinations are such that the number of  neutron-neutron, neutron-proton and proton-proton interactions cancel. In addition to having the correct number of  interactions, the single-particle energies and the residual interactions within each level, to a first approximation, cancel too \cite{Gar66}.

In order to understand the nature of the errors, in \cite{Fra03} a systematic study of nuclear masses was carried out using the shell model. This was achieved by employing
realistic Hamiltonians with a small random component. In
\cite{Hir04,Hir04b} we have analyzed in detail the error distribution for
the mass formulas of M\"oller et al.\cite{Moll95} and found a conspicuous long
range regularity that manifests itself as a double peak in the
distribution of mass differences \cite{Hir04}. This striking
non-Gaussian distribution was found to be robust under a variety
of criteria. By assuming a simple sinusoidal correlation, we could
empirically substract these correlations and made the average
deviation diminish by nearly $15\%$ \cite{Hir04b}.

In the present contribution we analyze the mass deviations
in the Finite Range Droplet Model (FRDM) of M\"oller et al. \cite{Moll95}, and
in the microscopically motivated mass formula of DZ \cite{Duf94},
and those obtained using the Garvey-Kelson relations \cite{Hir04b}.
The presence of strong correlations between mass errors in neighboring nuclei
is clearly exhibited, as well as the existence of a well defined chaotic signal in its 
power spectrum, when their correlations are analyzed as time series \cite{Hir04a,Bar04}. 
It is also shown that the intrinsic average mass error is smaller that 100 keV.

\section{Mapping the mass errors}
\label{sec:1}

\begin{figure}[h!]
\vskip -0.5cm
  \resizebox{0.50\textwidth}{!}{%
    \includegraphics{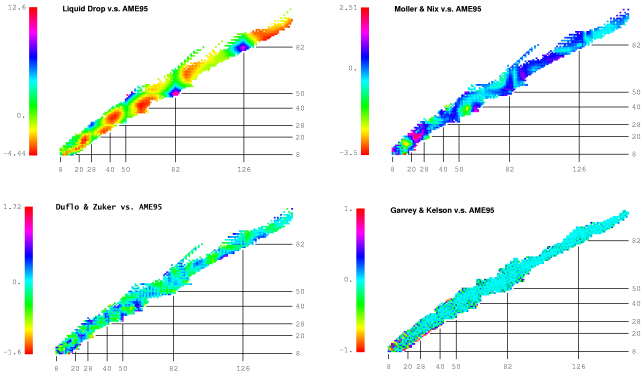} 
}
\caption{Mass differences from the  LDM, FRDM, DZ and our GK studies, in MeV, as functions of N and Z.}
\label{masas}
\end{figure}

In Fig. \ref{masas} we show a gray tone (color-coded on line) depiction of the distribution of mass deviations
in the Liquid Drop Model, in the FRDM, in the DZ calculations, and in GK calculations, 
in the proton number (N) - neutron number (Z) space. We 
can see large domains with a similar error (each  tone is
associated to the magnitude of the error). The shell closures are clearly seen in the LDM,
It is remarkable that very well defined correlated areas of the same gray tone exist
for the errors in the FRDM, and to a lesser extent in the DZ calculations, which
are a clear indication of remaining systematics and correlation.
In the GK calculations the errors are around 100 keV. Although the latter calculations
do not allow reliable extrapolations, they exhibit the calculability of nuclear masses
when enough local information (masses of neighbor nuclei or shell model realistic interactions) is available.

In order to measure and quantify the oscillatory patterns in the FRDM
observed in Fig. \ref{masas}, different cuts were performed
along selected directions on the N-Z plane. Given the large number
of chains which can be studied, we have selected those with the
largest number of nuclei with measured mass. For each cut a
Fourier analysis was performed, and the squared amplitudes are
plotted as a function of the frequencies on the right hand side of
each figure. 

\begin{figure}[h]
 \vskip -0.5cm
 \rotatebox{270}{  \resizebox{0.4\textwidth}{!}{%
   \includegraphics{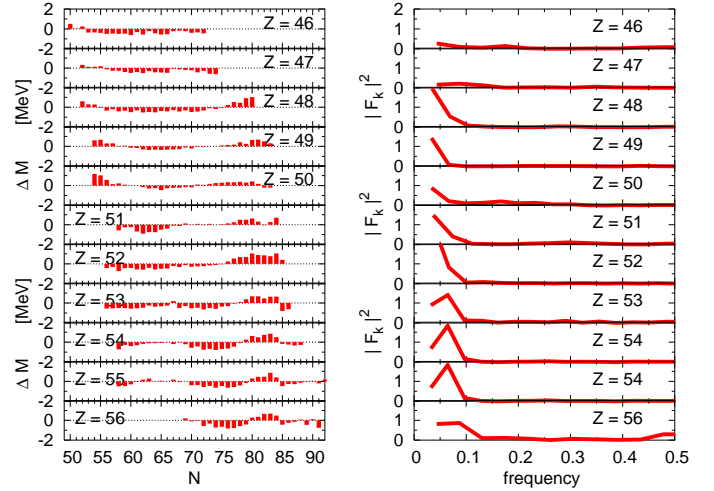}  }}  
\caption{Mass errors in the isotope chains Z=46 to 56, and their Fourier analysis.}
\label{z46}
\end{figure}

\begin{figure}[h]
\vskip -0.5cm
  \rotatebox{270}{  \resizebox{0.4\textwidth}{!}{%
   \includegraphics{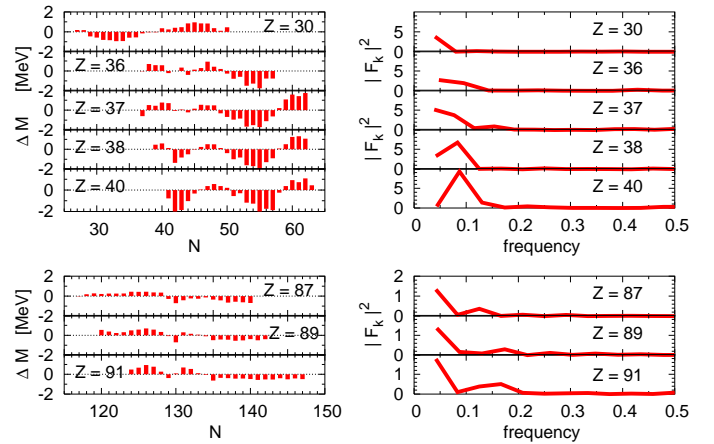}  }} 
\vskip -.9cm
\caption{Mass errors in the isotope chains with Z = 30,36, 37, 38, 40, 87, 89, 91,
and their Fourier analysis.}
\label{z30}
\end{figure}

We start our analysis for fixed N or Z, i.e. we selected different chains of isotopes or
isotones. Those isotopic chains with 20 or more nuclei with measured masses are
presented in Fig. \ref{z46} and \ref{z30}.
Until 1995, the element which had most
isotopes with measured masses was Cs (Z=55), with 34.

Fig. \ref{z46} displays the mass errors for the isotope chains Z =
46 to 56, and their Fourier analysis, nearly all exhibiting a
prominent peak around the low frequency $f \approx 1/20 = 0.05$.

Fig. \ref{z30} displays the mass errors for the isotope chains
Z = 30,36, 37, 38, 40, 87, 89, 91,  and
their Fourier analysis.

When the squared Fourier amplitudes are plotted as functions of the frequency $\omega=k/N$
using a log-log scale, the corresponding spectral distributions can
then be fitted to a power law of the form
$|F(\omega)|^2\sim\omega^m$.
For the 18 chains listed, the fitted slopes $m$ are
\begin{equation}
m^{(1)}_{FRDM} = -1.18 \pm  0.17, \; ~~~m^{(1)}_{DZ} = -0.67 \pm 0.16.
\end{equation}
They give values close to -1.2  in the FRDM data and around -0.7 for the deviations found by DZ.
The former is consistent with a
frequency dependence of  $f^{-1}$ characteristic of quantum chaos \cite{Rel02}, while the 
latter suggest a tendency towards a more random behavior characteristic of white noise.

\section{The boustrophedon line}
\label{sec:2}
\begin{figure}[h]  
 \vskip -0.5cm
\resizebox{0.4\textwidth}{!}{
   \includegraphics{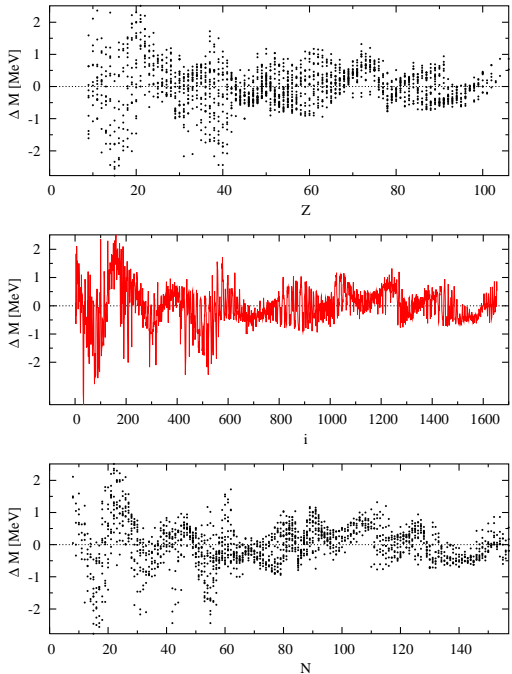}
  }
\caption{Mass differences plotted as function of Z (top), N (bottom), and of an ordered list
(middle)} 
\label{difmas-l}
\end{figure}

Plotting the mass differences for different Z, Fig. \ref{difmas-l} top, and
for different N,  Fig. \ref{difmas-l} bottom, is very common in mass calculations.
Both plots exhibit some degree of structure. In this way we obtain a plot of mass 
differences as a function of Z, with all the isotopes
plotted along the same vertical line, see Fig. \ref{difmas-l}.
The difficulty in quantifying these
regularities lies in the simple fact that the are many nuclei with a given N or Z.
For this reason in \cite{Hir04} we have analyzed the data using different cuts.
Another way to organize the FRDM mass errors for the 1654 nuclei with measured masses
is to order them in a single list, numbered in increasing order. To avoid jumps,
we have ordered the isotopes along a $\beta{}o\upsilon\tau\rho{}o\phi\eta\delta\acute{o}\nu$ 
(boustrophedon) line \cite{Hir04b}, which literally means "in the way the ox ploughs".
Nuclei were ordered in increasing mass order. For a given even A, they were
accommodated following the increase in N-Z, and those
nuclei with odd A starting from the largest value of N-Z, and going on in decreasing order.
 The middle panel
exhibits the same mass differences plotted against the order number, from 1 to 1654,
providing an univalued function,

The presence of strong correlations in the M\"oller at al mass
differences is apparent from the plot. Regions with large positive
or negative errors are clearly seen. In contrast, the distribution of errors for the data of Duflo and Zuker (not shown, see Ref. \cite{Hir04a}) is closer to the horizontal axis,
and the correlations are less pronounced, although not completely
absent.

\begin{figure}[h!]
 \vskip -0.5cm
\rotatebox{270}{  \resizebox{0.4\textwidth}{!}{%
   \includegraphics{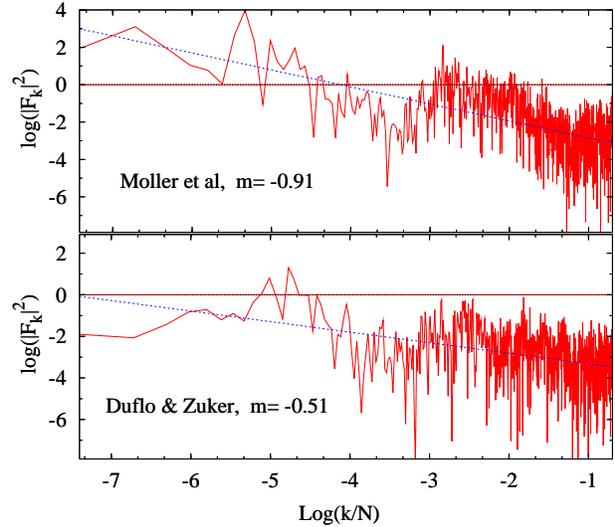}  }}
\caption{Log-log plot of the squared amplitudes of
the Fourier transforms of the mass differences, as functions of
the order parameter(top). Data from FRDM (top) and from
Duflo and Zucker (bottom).} \label{fourier-l}
\end{figure}

The ordering provides a single-valued function,
whose Fourier transform can be calculated. The squared amplitudes are presented
in Fig. \ref{fourier-l}.
The slopes are 
\begin{equation}
m^{(2)}_{FRDM} = -0.91 \pm  0.05, \; ~~~m^{(2)}_{DZ} = -0.51 \pm 0.05,
\end{equation}
 for the FRDM and DZ mass differences. 

While this ordering is quite different from the chains along N and Z, the slopes are very similar.

To understand the possible origin of these spectral distributions,
it is worth recalling that, while the FRDM calculations
involve a liquid droplet model plus mean field corrections,
including deformed single particle energies through the Strutinsky
method and pairing \cite{Moll95}, the DZ calculations depend on
the number of valence proton and neutron particles and holes,
including quadratic terms motivated by the microscopic
Hamiltonian \cite{Duf94}.

\section{Local analysis of the differences between measured and calculated masses}

We apply the GK procedure to all nuclei in the 2003 compilation \cite{Aud03} where at least one of the relations 
 \begin{eqnarray}
\lefteqn{
-M(N\!+\!1,Z\!-\!2)\!+\!M(N\!+\!1,Z)\!-\!M(N\!+\!2,Z\!-\!1)}
\label{GK1}\\
&&+M(N\!+\!2,Z\!-\!2)\!-\!M(N,Z)\!+\!M(N,Z\!-\!1)=0,
\nonumber\\
\lefteqn{
M(N\!+\!2,Z)\!-\!M(N,Z\!-\!2)\!+\!M(N\!+\!1,Z\!-\!2)}
\label{GK2}\\
&&-M(N\!+\!2,Z\!-\!1)\!+\!M(N,Z\!-\!1)\!-\!M(N\!+\!1,Z)=0.
\nonumber
\end{eqnarray}
is applicable.

These simple equations are based on the independent-particle shell model
and, furthermore, constructed such that neutron-neutron, neutron-proton,
and proton-proton interactions cancel.
Both GK relations provide an estimate for the mass of a given nucleus
in terms of five of its neighbors.
This calculation can be done in six different forms, as we can choose any of the six terms in the formula to be evaluated from the others.
Using both formulas, we can have a maximum of 12 estimates for the mass of a given nucleus,
if the masses of all the required neighboring nuclei are known.
Of course, there are cases where only 11 evaluations are possible, and so on.
About half of all nuclei with measured masses~\cite{Aud03}
can be estimated in 12 different ways
and, in all cases, our estimate corresponds to the average value.
To our knowledge, the systematic application of GK relations in this extended fashion \cite{Bar04} is new. 
 
Using the GK procedure we obtain a very specific prediction, determined by that of its neighbors.  In this procedure there are no free parameters and there is no fit to the data,  just a prediction of nuclear masses arising from those of  its neighbors. 
In what follows we compare the mass deviations found in three of the global methods (LDM, FRDM, DZ) and our GK studies. The  corresponding $\sigma_{r.m.s.}$ deviations,
defined as
\begin{equation}
\sigma_{r.m.s.} = \left[ {\frac 1 N} \sum_{i=1}^N (M^i_{exp} - M^i_{th})^2 
\right]^{1/2}
\end{equation}
are displayed in Table \ref{rms},
\begin{table}
\begin{tabular}{cccccc}
~~\\
model                                 &LDM   &FRDM  &DZ      &GK \\
$\sigma_{r.m.s.}$ &3447    & 669    & 346  & 189 \\~~\\
GK relations   & 1-12     & 4-12  &  7-12  & 10-12  & 12 \\
A $\ge$ 16      &   189  & 162 & 117  & 95 & 86 \\
A $\ge$ 60       &  123   & 102  & 87   &  81  & 80
\end{tabular}
\caption{$\sigma_{r.m.s.}$ mass differences, in keV for the LDM, FRDM, DZ and GK calculations, and for different GK calculations.} \label{rms}
\end{table}
where we also include the smaller samples GK-n which involve the application of n or more GK relations, for which the average deviation is also quoted. Note the systematic decrease in the errors as a consequence of a better determination of the masses, proportional to the number of GK relations applied, 
for each of the four methods employed. In our best scenario, that of GK-12, we find an r.m.s. deviation of 80 keV, almost an order of magnitude smaller than the FDRM  one.

\section{Conclusions}
\label{conc}

In summary, a careful use of several global mass formulas
and a systematic application of the Garvey-Kelson relations
imply that there is no evidence
that nuclear masses cannot be calculated
with an average accuracy of better than 100~keV.
While mass errors in mean-field calculations like the FRDM
behave in a manner akin to quantum chaos,
with a slope in the power spectrum close to $-1$,
microscopic models' results correspond to smaller slopes.
Finally, for the local GK relations
the remaining mass deviations
behave very much like white noise.
These results seem to confirm
that the chaotic behavior in the fluctuations
arises from neglected many-body effects.

In other words, the chaoticity discussed in
\cite{Boh02}, according to the criteria  put forward in \cite{Rel02}, seems indeed to be 
present in the deviations induced by calculations using the M\"oller et al. liquid droplet mass
formula, while it tends to diminish in the microscopically motivated
calculations of Duflo and Zucker. 
 While for the liquid droplet model plus shell
corrections a quantum chaotic behavior  $m \approx
1$ is found, errors in the microscopic mass formula have   
$m \approx 0.5 $, closer to white noise.
Given that both models attempt
to describe the same set of experimental masses, our analysis
suggests that quantum fluctuations in the mass differences arising
from substraction of the regular behavior provided by the liquid droplet model 
plus shell corrections, may have their origin in an incomplete consideration
of many body quantum correlations, which are partially included in
the calculations of Duflo and Zuker.


%
%

\end{document}